\documentstyle[12pt,leqno]{article}
\begin{document}

\begin{center}
{ \normalsize ANOTHER STATE ENTANGLEMENT MEASURE}\\
$\;$ \\

{\normalsize Oscar A. Nagel\footnote{Depto. de F\'{\i}sica, Universidad Nacional del Sur, 
Bahia Blanca, Argentina. e-mail: nagel@uns.edu.ar},
\normalsize and Guido A. Raggio\footnote{FaMAF-Universidad Nacional de C\'ordoba, C\'ordoba, Argentina. e-mail: raggio@famaf.unc.edu.ar}}
\end{center}

Given a state $\omega $ of the (minimal $C^*$-) tensor product $A \otimes B$ of unital $C^*$-algebras
$A$ and $B$,
its marginals are the states of $A$ and $B$ defined by
\[ \omega^A (a) = \omega (a \otimes {\bf 1}_B)\;,\;\; a \in A \;, \;\;
\omega^B (b) = \omega ( {\bf 1}_A \otimes b) \;,\;\; b \in B\;.\]
Given a state $\rho$ of $A$ and a state $\phi$ of $B$, there is a unique state
$ \omega $ of $A \otimes B$ such that $\omega ( a\otimes b) = \rho (a) \phi (b)$ for
all $a\in A$ and all $b \in B$; we denote this state by $\rho \otimes \phi$.
A {\bf product-state} of $A \otimes B$ is a state $\omega$ of $A\otimes B$ such that
$\omega = \omega^A \otimes \omega^B$. 
We write $S_{\pi}(A\otimes B)$ for the product-states of $A \otimes B$. The convex hull of 
$S_{\pi}$, written $co(S_{\pi}(A \otimes B))$, is the set of finite convex
combinations of product states. The states of $A\otimes B$ in the norm-closure 
of $co(S_{\pi}(A \otimes B))$ are usually identified with the {\bf separable} states of the
composite system whose observables are described by $A \otimes B$; the states which are not separable are termed {\bf entangled}.\\

For a state $\omega$ of a unital $C^*$-algebra $A$, consider its finite convex
decompositions: $\omega = \sum_{j=1}^n \lambda_j \omega_j$, with $0 \leq \lambda_j\leq 1$, 
$\sum_{j=1}^n \lambda_j=1$, and $\omega_j$ a state of $A$. Such a decompositon will be written
$[\lambda_j, \omega_j]$ and ${\cal D}_{\omega}$  denotes all such finite convex decompositions.\\

Consider the realtive entropy $( \rho , \phi ) \to S( \rho, \phi )$ for pairs of states $\rho$
and $\phi$ of a
unital $C^*$-algebra. We use the original convention of Araki [1]\footnote{If $A$ is the algebra of
bounded linear operators on a Hilbert space, then $S(\rho, \phi)=Tr(D_{\rho}(\log (D_{\rho}-\log (D_{\phi})))$, for normal states,
where $D_{\rho}$ (resp. $D_{\phi}$) is the density operator for which $\rho (a)=Tr(D_{\rho}a)$, $a \in A$.}, which is also that used in [2]
which we use as a standard reference for the
 properties of relative entropy. We propose the following measure of entanglement
 \begin{equation}
 E ( \omega ) = \inf_{[\lambda_j, \omega_j]\in {\cal D}_{\omega} }
 \sum_{j=1}^n \lambda_j S( \omega_j, \omega_j^A\otimes \omega_j^B) \;.\label{E}\\
 \end{equation}

 We say a map $\alpha$ from $A\otimes B$ into $C \otimes D$
{\bf commutes with marginalization} if  for every state $\omega$ of
$C\otimes D$  one has $(\omega \circ \alpha)^A\otimes (\omega
\circ \alpha)^B =(\omega^C \otimes \omega^D) \circ \alpha$.\\

We have the following result, whose proof will be provided in a forthcoming paper [3], along
with result about a class of entanglement measures akin to (\ref{E}):

\begin{enumerate}
\item  $0 \leq E( \omega ) \leq S( \omega, \omega^A\otimes \omega^B)$ with equality in the right-hand side
inequality if $\omega$ is a pure state. $E(\omega)=0$ if $\omega$ is a product-state.
\item $ E( \cdot )$ is convex (and in general not affine).
\item If $\alpha$ and $\beta$ are, respectively, *-isomorphisms of $A$ onto $C$ and of $B$ onto $D$ 
($A,B,C$ and $D$ are unital $C^*$-algebras)
then $E ( \omega \circ ( \alpha \otimes \beta ) )=E ( \omega ) $ for every state of
$C \otimes D$.
\item If $\gamma :A \otimes B \to C\otimes D$ is a unital, linear, continuous, Schwarz-positive map
($\gamma (z^*z)\geq \gamma (z)^*\gamma (z)$ for every $z \in A \otimes B$)
 which commutes with marginalization, then $E( \omega \circ \gamma ) \leq E ( \omega )$ for
every state $\omega$ of $C\otimes D$.
\item If $\omega$ is separable then $E( \omega )=0$.
\item $E( \omega )=0$ iff $\omega$ lies in the $w^*$-closure of $co(S_{\pi}(A \otimes B))$.
\item For $n$ ($n \geq 1$) states $\omega_1, \omega_2, \cdots ,\omega_n$ of $A \otimes B$,
\begin{equation}  E ( (\omega_1\otimes \omega_2 \otimes \cdots \otimes \omega_n)
 \circ \zeta_n ) \leq \sum_{j=1}^n E ( \omega_j)\;,\label{subadd}\end{equation}
where $\zeta_n$ is the *-isomorphism
\begin{equation}  \left\{\underbrace{A \otimes A \otimes \cdots \otimes A}_{n}\right\} \otimes \left\{
\underbrace{B \otimes B \otimes \cdots \otimes B}_{n}\right\}
\stackrel{\zeta_n\;\;}{\rightarrow}
\underbrace{(A\otimes B) \otimes (A \otimes B) \otimes \cdots (A \otimes B)}_{n},\label{iso}
\end{equation}
given by $\zeta_n ( (a_1\otimes a_2 \otimes \cdots \otimes a_n)\otimes (b_1\otimes b_2\otimes \cdots \otimes b_n)) =
(a_1\otimes b_1)\otimes (a_2\otimes b_2) \otimes \cdots (a_n \otimes b_n)$. 
One has,
\begin{equation} \lim_{n \to \infty} n^{-1}E((\omega\otimes \omega \otimes \cdots \otimes \omega)
 \circ \zeta_n ) \leq  E ( \omega)\;.\label{ext}\end{equation}
 In both (\ref{subadd}) and (\ref{ext}), the left-hand side is computed with respect to marginalization
  with respect to the two factors in $\{ \,\}$-brackets in (\ref{iso}).
  \item If $A$ or $B$ is abelian then $E\equiv 0$.
\item Let ${\cal M}_{\omega} $ be the (Radon)-measures on the state space with barycenter $\omega$,
 then 
 \[  E( \omega )=  \inf_{\{\mu \in {\cal M}_{\omega}\}}\int \mu (d \phi ) S( \phi , \phi^A\otimes \phi^B)\;,\]
  and there
exists $\mu_o \in {\cal M}_{\omega}$ such that
\[ E ( \omega ) = \int \mu_o ( d \phi ) S( \phi, \phi^A \otimes \phi^B ) \;. \]
\end{enumerate}

\begin{center} -----\end{center}

The crucial condition of ``commmutation with marginalization'' involved in property 4. of $E$ is 
met by the ``LQCC'' maps considered in [4]. ``LQCC'' means ``local quantum operations'' with 
``classical communication'', and
these are the relevant maps in the games that Alice and Bob play.\\

Like most known entanglement measures (see e.g., [4,5]), except that devised by Vidal and Werner [6], the 
calculation of $E$ involves an infimum over a rather unmanageable set. Using Kosaki's  variational
expression ([7]) for the relative entropy, one obtains a lower bound on $E$ which can be possibly used to 
devise a strategy to show that $E( \omega )>0$ for a specific state $\omega$.\\

One can replace the relative entropy in the definition of $E$ by other, suitable functions, e.g.
$\parallel \phi - \phi^A\otimes \phi^B\parallel$, without losing the basic properties of $E$;
this is studied in [3].\\

In a previous version of this announcement [8], we claimed that $E$ was additive, that is, 
equality holds in (\ref{subadd}). We withdraw this claim because we have found a mistake in
our ``proof''. \\

\noindent REFERENCES

\begin{enumerate}
\item H. Araki: Relative entropy for states of von Neumann algebras I [\& II]. Publ. RIMS Kyoto Univ. 11, 809-833 (1976);
 [\& 13: 173-192 (1977)].
\item M. Ohya, and D. Petz: Quantum Entropy and Its Use. Springer-Verlag Berlin Heidelberg 1993.
\item O.A. Nagel, and G.A. Raggio: A family of state entanglement measures. Work in progress.
\item M.J. Donald, M. Horodecki, and O. Rudolph: The uniqueness theorem for entanglement measures.
 J. Math. Phys. 43, 4252-4272 (2002).
\item V. Vedral: The role of relative entropy in quantum information theory. Rev. Mod. Phys. 74, 197-234 (2002).
\item G. Vidal, and R.F. Werner: A computable measure of entanglement. Phys. Rev. A 65, 032314 (2002).
\item H. Kosaki: Relative entropy for states: a variational expression. J. Operator Th. 16, 335-348 (1986).
\item O.A. Nagel and G.A. Raggio: Another state entanglement measure. quant-ph/0306024 v2.
\end{enumerate}

\end{document}